\documentstyle[preprint,aps,tighten]{revtex}
\newcommand{\Tr}{{\rm Tr}\,}
\newcommand{\Det}{{\rm Det}\,}
\newcommand{\ie}{{\it i.e.\/}}
\newcommand{\one}{\mbox{{\sf 1}\hspace{-0.20em}{\rm l}}}
\newcommand{\beq}{\begin{equation}}
\newcommand{\eeq}{\end{equation}}
\newcommand{\bara}{\begin{eqnarray}}
\newcommand{\ear}{\end{eqnarray}}

\newcommand{\cH}{{\cal H}_}

\begin{document}

\title{The Parity Bit in Quantum Cryptography}

\author{Charles H. Bennett$^{(1)}$, Tal Mor$^{(2)}$ and 
John A. Smolin$^{(3)}$}

\address{(1)IBM Research Division, Yorktown Heights, NY 10598; 
(2) Physics Department, Technion, Israel;
(3) Physics Department, University of California at Los Angeles, 
Los Angeles, CA 90024;}

\date{\today}
\maketitle

\begin{abstract}

An $n$-bit string is encoded as a sequence of 
non-orthogonal quantum states. The parity bit of that
$n$-bit string is described by one of  two density matrices, 
$\rho_0^{(n)}$
and $\rho_1^{(n)}$,  
both in a Hilbert space of dimension $2^n$.  In order to
derive the parity bit the receiver must distinguish between the  two
density matrices, e.g., in terms of optimal mutual information. In 
this paper we find the measurement which provides  the optimal mutual
information about the parity bit  and calculate that information.  We
prove that this information decreases exponentially with the length 
of the string in the case where the single bit states are almost 
fully overlapping. We believe this result will be useful in proving
the ultimate security of quantum crytography in the presence of 
noise.
\end{abstract}


\section{Introduction}

A major question in quantum information theory 
\cite{Hol,Davies,Helstrom,Per93,Levitin,Fuchs}
is ``how well can two
quantum states, or more generally, two density matrices $\rho_0$ 
and $\rho_1$, be distinguished?''
In terms of a communication scheme this question is 
translated to an identification task: 
A sender (Alice) sends a bit $b=i$ ($i=0;1$) 
to the receiver (Bob) by sending the 
quantum state $\rho_i$, and the receiver does his best to identify
the value of that bit, {\em i.e.}~the quantum state.
The Two-dimensional Hilbert space $\cH2$ is usually used to 
implement such a binary channel, so the transmitted signals can be 
polarization states 
of photons, spin-states of spin-half particles {\em etc}.
The transmitted states may be pure states or density
matrices, and need not be orthogonal.
Usually, the mutual information $I$ is used to describe 
distinguishability, 
such that $I=0$ means indistinguishable, and $I=1$ 
(for a binary channel) means perfect distinguishability.
The ensemble of signals is agreed on in advance, and  
the main aim of Alice and Bob is to optimize the avarage mutual 
information over the different possible measurements at the 
receiving end. For a (simple) example, 
two orthogonal pure states transmitted through an error-free
channel are perfectly distinguishable; The optimal mutual 
information ($I=1$) 
is obtained if Bob measures in an appropriate basis.
Finding the optimal mutual information is still an open 
question for most ensembles.
Some cases with known analytic solutions are the case of two pure 
states and the case of two density matrices in two dimensions with
equal determinants\cite{Levitin,Fuchs}. 
There are no known analytic solutions for two non-trivial density
matrices in dimensions higher than two.
In this paper we find a solvable case which has also practical
implications.

Let a single bit be presented by one of two quantum 
states, $\rho_0$ and $\rho_1$, in $\cH{2}$. 
These can be either pure states
or density matrices with equal determinants.
Suppose Bob wants to learn 
the parity bit (exclusive-OR) of an $n$-bit string of these
bits and not the specific value of each bit.
The parity bit 
is described by one of two density matrices 
$\rho_0^{(n)} $
and  
$\rho_1^{(n)} $ which lie in a $2^n$-dimensional Hilbert 
space $\cH{2^n}$.  We solve the case of the distinguishablity of 
these parity density matrices when the two possible signals 
(of each single bit) are sent with equal probabilities.
Let $x$ be any classical string of $n$ such bits, and  
$\rho_x = \rho_{(1^{st} \ {\rm bit})} \ldots  
          \rho_{(n^{th} \ {\rm bit})}$  
be the density
matrix made up of the tensor product of the signaling 
states $\rho_(i)$ corresposding to the i$^{th}$ bit of $x$.
Formally, we distinguish between the two density matrices:
\begin{equation}
\rho_0^{(n)}=\frac{1}{2^{n-1}}\sum_{x\,|\,p(x)=0}\!\!\!\!\!\rho_x\ 
\quad {\rm and} \quad
\ \rho_1^{(n)}=\frac{1}{2^{n-1}}\sum_{x\,|\,p(x)=1}\!\!
\!\!\!\rho_x \ ,
\end{equation}
where the sum is over all possible strings with the same parity
(each sent with equal probability $(1/2^n)$ 
and $p(x)$ is the parity function of $x$.
We show a simple way to write the parity density matrices 
and we calculate the optimal mutual information which can be 
obtained.

Parity bits are often used in quantum 
cryptography\cite{BB84,Eke91,Ben92,BCJL},
where they play a crucial role in error-correction and privacy 
amplification\cite{BBBSS,BBR,BBCM}. The question of 
security of quantum cryptography is
yet open, and our 
results may have several implications for attacking this issue.
In particular, the special case where the two
signaling states have large overlap is important in the analysis
of the security of quantum key distribution against powerful 
multi-particle eavesdropping attacks.

In Section \ref{sec2} we find a simple way to write the density 
matrices of the parity bit for any $n$; We show that they can be 
put in a block diagonal form and we explain the importance of 
that fact. In Section \ref{sec3} we investigate the 
distinguishability of the parity
matrices; The optimal
measurement which distinguishes them is found to be a standard
(von Neumann) measurement in an entangled
basis (which is a generalization of the Bell basis of two particles);
We calculate exactly 
the optimal mutual information which is derived
on the parity bit by performing that optimal measurement. 
In Section \ref{sec4} we show that for two
almost fully overlapping states, the optimal mutual information $I_M$
decreases exponentially with the length of the string;  
This is the main result of our paper due to its possible 
importance to quantum cryptography.
While exponentially small, this optimal information is nevertheless
considerably greater than the information that would have been 
obtained by measuring each bit separately and classically combining 
the results of these measurements, thus, we prove the advantage of
such ``joint'' measurements.
Going back to the parity matrices obtained in Section \ref{sec2}
we are also able to calculate the maximal deterministic (conclusive) 
information;
This is done in Section \ref{sec5} where we also confirm a result 
previously obtained by Huttner and Peres \cite{HP} for two bits.
In Section \ref{sec6} we repeat the calculation
of the optimal mutual information for the more general case where 
the bits are represented by non-pure states (in $\cH2$ and 
with equal determinants).
In Section \ref{secd} 
we briefly discuss the implications of our 
results to the security of quantum cryptography. 

\section{Density Matrices for Parity Bits}\label{sec2}

Let Alice send $n$ bits.  The possible values of a single bit
($0$ or $1$) are represented by 
\beq   
\psi_0={\cos\alpha\choose\sin\alpha} \quad \ {\rm and } \ \quad 
\psi_1={\cos\alpha\choose -\sin\alpha} \label{pure-state}\eeq
respectively.
In terms of density matrices these are:
\beq \rho_0^{(1)} = 
  \left(\begin{array}{rr}   c^2 &  sc  \\  sc &  s^2     
  \end{array}\right) 
\quad \quad {\rm and } \quad \quad 
 \rho_1^{(1)} = 
  \left(\begin{array}{cc}   c^2 &  -sc  \\  -sc &  s^2     
  \end{array}\right),\label{rho^1} \eeq
where we use a shorter notation 
$s \equiv \sin\alpha$; $c \equiv \cos\alpha$,
for convenience, and the superscript $[]^{(1)}$ is explained in the 
following paragraph.

The parity bit of an $n$-bit string is the exclusive-OR of all 
the bits in the string. In other words, the parity is 1 if there 
are an odd number of
1's and 0 if there are an even number.
The parity density matrices of $n$ bits will be denoted as
$\rho_0^{(n)}$ and $\rho_1^{(n)}$ in case the parity is 
`0' and `1' respectively.
Using these density matrices we define also the {\em total} density 
matrix
$\rho^{(n)} \equiv  \frac{1}{2} (\rho_0^{(n)} + \rho_1^{(n)})$ 
and the {\em difference} density matrix
$\Delta^{(n)} \equiv \frac{1}{2} (\rho_0^{(n)} - \rho_1^{(n)})$, so 
that 
\beq \rho_0^{(n)} = \rho^{(n)} + \Delta^{(n)} 
 \quad \quad {\rm and } \quad \quad 
\rho_1^{(n)} = \rho^{(n)} - \Delta^{(n)} \ . \label{rho^n} \eeq
 
The one-particle density matrices 
(equation \ref{rho^1}) also describe the parities of one 
particle, 
and therefore we can calculate 
\beq \rho^{(1)} = \frac{1}{2} (\rho_0^{(1)} + \rho_1^{(1)}) = 
  \left(\begin{array}{cc}   c^2 &  0  \\  0 &  s^2     
  \end{array}\right),\label{rho} \eeq
\beq \Delta^{(1)} = \frac{1}{2} (\rho_0^{(1)} - \rho_1^{(1)}) = 
  \left(\begin{array}{cc}   0 &  sc  \\  sc &  0     
  \end{array}\right) \label{Delta} \ . \eeq
The density matrices of the parity bit
of two particles are:
\bara \rho_0^{(2)} &=& \frac{1}{2} (\rho_0^{(1)} \rho_0^{(1)} + 
                     \rho_1^{(1)} \rho_1^{(1)})  \nonumber \\ 
     \rho_1^{(2)} &=& \frac{1}{2} (\rho_0^{(1)} \rho_1^{(1)} + 
                             \rho_1^{(1)} \rho_0^{(1)})  \ear
where the multiplication is a tensor product.
The total density matrix is 
\bara 
\rho^{(2)} &=& \frac{1}{2} (\rho_0^{(2)} + \rho_1^{(2)}) \nonumber \\
      &=& \frac{1}{4} [\rho_0^{(1)} (\rho_0^{(1)} + \rho_1^{(1)})
       + \rho_1^{(1)} (\rho_1^{(1)} + \rho_0^{(1)})] \nonumber \\
      &=& \rho^{(1)} \rho^{(1)} \ , \nonumber \ear 
which, by using the basis  
\bara |b_0\rangle \equiv {1 \choose 0}_1 
      {1 \choose 0}_2 = \left(  \begin{array}{c} 
                    1 \\ 0 \\ 0 \\ 0 \end{array} \right)
  \ ; \quad  |b_1\rangle \equiv 
      {1 \choose 0}_1 
      {0 \choose 1}_2 = \left(  \begin{array}{c} 
                    0 \\ 1 \\ 0 \\ 0 \end{array} \right)
  \ ; \nonumber \\ |b_2\rangle \equiv
      {0 \choose 1}_1 
      {1 \choose 0}_2 = \left(  \begin{array}{c} 
                    0 \\ 0 \\ 1 \\ 0 \end{array} \right)
  \quad   {\rm and} \quad |b_3\rangle \equiv  
      {0 \choose 1}_1 
      {0 \choose 1}_2 = \left(  \begin{array}{c} 
                    0 \\ 0 \\ 0 \\ 1 \end{array} \right)
\label{basis}
\ear   
in $\cH4$, can be written as 
\bara \rho^{(2)} = 
       \rho^{(1)} \rho^{(1)} =  
  \left(\begin{array}{cccc}    c^4   &    0    &    0    &    0   \\ 
                                0    & c^2 s^2 &    0    &    0   \\ 
                                0    &    0    & c^2 s^2 &    0   \\ 
                                0    &    0    &    0    &   s^4  \\ 
  \end{array}\right). \ear
The difference density matrix is
\bara 
\Delta^{(2)} &=& \frac{1}{2} 
(\rho_0^{(2)} - \rho_1^{(2)}) \nonumber \\
      &=& \frac{1}{4} [\rho_0^{(1)} (\rho_0^{(1)} - \rho_1^{(1)})
 + \rho_1^{(1)} (\rho_1^{(1)} - \rho_0^{(1)}) ] \nonumber \\
    &=& \Delta^{(1)} \Delta^{(1)} = 
\left(\begin{array}{cccc}     0    &    0    &    0    & c^2 s^2 \\ 
                              0    &    0    & c^2 s^2 &    0    \\ 
                              0    & c^2 s^2 &    0    &    0    \\ 
                           c^2 s^2 &    0    &    0    &    0    \\ 
\end{array} \right) \ . \ear 

The density matrices of the parity bit
of $n$ particles can be written recursively:
\bara \rho_0^{(n)} = \frac{1}{2} (\rho_0^{(1)} \rho_0^{(n-1)} + 
                          \rho_1^{(1)} \rho_1^{(n-1)})  \nonumber \\
      \rho_1^{(n)} = \frac{1}{2} (\rho_0^{(1)} \rho_1^{(n-1)} + 
                             \rho_1^{(1)} \rho_0^{(n-1)}) \ , \ear
leading to 
\beq \rho^{(n)} = \frac{1}{2} (\rho_0^{(n)} + \rho_1^{(n)}) 
      = \rho^{(1)} \rho^{(n-1)}   \ , \nonumber  \eeq
and
\beq \Delta^{(n)} = \frac{1}{2} (\rho_0^{(n)} - \rho_1^{(n)})
      = \Delta^{(1)}  \Delta^{(n-1)}   \ .\nonumber   \eeq
Using these expressions recursively we get
\beq \rho^{(n)} = 
       (\rho^{(1)})^n   \label{rho-n} \eeq
which is diagonal, and
\beq \Delta^{(n)} = 
      (\Delta^{(1)})^n      \label{Delta-n} \eeq
which has non-zero terms only in the secondary diagonal.
The density matrices  
$  \rho_0^{(n)}$
and $\rho_1^{(n)}$ are now immediately derived for any $n$
using equation (\ref{rho^n}):
\bara \rho_0^{(n)} = 
       (\rho^{(1)})^n +
      (\Delta^{(1)})^n      \nonumber  \ear
and
\bara 
     \rho_1^{(n)} = 
       (\rho^{(1)})^n -
      (\Delta^{(1)})^n    \ . \label{rho_p^n} \ear
As an illustrative example we write $\rho_0$ and $\rho_1$ for two
particles: 
\bara \rho_0^{(2)} = 
 \left(\begin{array}{cccc}    c^4   &    0    &    0    & c^2 s^2 \\ 
                               0    & c^2 s^2 & c^2 s^2 &    0    \\ 
                               0    & c^2 s^2 & c^2 s^2 &    0    \\ 
                            c^2 s^2 &    0    &    0    &   s^4   \\
  \end{array}\right) \ ; 
      \rho_1^{(2)} = 
 \left(\begin{array}{cccc}    c^4   &    0    &    0    &-c^2 s^2 \\ 
                               0    & c^2 s^2 &-c^2 s^2 &    0    \\ 
                               0    &-c^2 s^2 & c^2 s^2 &    0    \\ 
                           -c^2 s^2 &    0    &    0    &   s^4   \\
  \end{array}\right) \ . 
\ear
The only non-zero terms in the parity density matrices are the 
terms in the diagonals for any $n$, thus the parity density matrices
have an X-shape in that basis.

\medbreak

The basis vectors can be reorganized 
to yield block-diagonal matrices built of $2\times 2$  blocks. 
The original basis vectors (see, for example, equation~\ref{basis}), 
$|b_i\rangle$, are simply $2^n$-vectors where 
the $i^{\rm th}$ element of 
the $i^{\rm th}$ basis vector 
is $1$ and all other elements are $0$ ($i$ ranges from $0$ 
to $2^n-1$).
The new basis vectors are related to the old as follows:  
\bara
|b'_i\rangle = |b_{i/2}\rangle\ {\rm for\ even}\ i \quad
{\rm and}\ |b'_i\rangle=|b_{2^n-(i+1)/2}\rangle\ {\rm for\ odd}\ i
\ \ .\ear
The parity density matrices are now, in the new basis (we omit 
the $'$ from
now on as we will never write the matrices in the original basis):
\beq \rho_p^{(n)} = 
      \left(  \begin{array}{cccc} 
         B_p^{[j=1]} & 0 & \ldots & 0 \\
           0 & B_p^{[j=2]} & \ldots & 0 \\
           0 & 0 & \ldots & B_p^{[j=2^{(n-1)}]} 
                                                \end{array} \right)
\label{block-diag} \eeq
where the subscript $p$ stands for the parity ($0$ or $1$). Each of 
the 2$\times$2 matrices has the form 
\beq   B_p^{[j]} = \left(\begin{array}{cc}   
         c^{2(n-k)} s^{2k}    &  \pm c^n s^n             \\ 
         \pm c^n s^n          &  c^{2k} s^{2(n-k)} 
  \end{array}\right), \label{Bpj} \eeq
with the plus sign for $p=0$ and the minus sign for $p=1$, and
$0 \le k \le n$, 
and all these density matrices satisfy
$ \Det B_p^{[j]} = 0 $. 
The first block ($j=1$) has $k=0$; there are 
${n \choose 1}$ blocks which have  
$k=1$ or $k=n-1$; there are ${n \choose 2}$ $j$'s which 
have $k=2$ or $k=n-2$, etc.  
This continues until $k=(n-1)/2$ for odd $n$.  For even $n$ the 
process continues up to $k=n/2$ with the minor adjustment that 
there are 
only $\frac{1}{2}{n \choose n/2}$ $j$'s of $k=n/2$.  This enumeration
groups blocks which are identical or identical after interchange
of $k$ and $n-k$ and accounts for all $2^n/2$ blocks.  We will see 
later that blocks identical under interchange of  $k$ and $n-k$ will
contribute the same mutual information about the parity bit, thus 
we have grouped them together.

With the density matrices written in such a block-diagonal form 
of 2x2 blocks the problem of finding the optimal mutual information
can be analytically solved.  It separates into two parts:
\begin{itemize}
\item Determining in which of $2^n/2$ orthogonal 2d subspaces
(each corresponding to one of the 2x2 blocks) the system lies. 
\item Performing the optimal measurement within that subspace.
\end{itemize}
The subspaces may be thought of as $2^n/2$ parallel channels,
one of which is probabilistically chosen and used to encode the 
parity
by means of a choice between two equiprobable pure states within that
subspace (these two states are pure because the $B_0$ and $B_1$ 
matrices each
have zero determinant).  We shall present in the next section the 
optimal
measurement that yields 
the optimal mutual information transmissible
through such a two-pure-state quantum channel. The channel then 
corresponds to a classical binary symmetric channel
(BSC), \ie~a classical one-bit-in one-bit-out channel whose output
differs from its input with some error probability $p_j$ independent
of whether the input was 0 or 1.  
The optimal mutual information in each subchannel is the optimal
mutual information
of a BSC with error
probability $p_j$ and
is $I_2(p_j)=1-H(p_j)$,  
with $H(x)=-x\log_2x-(1-x)\log_2(1-x)$,    
the Shannon
entropy function. 
The optimal mutual information $I_M$ for
distinguishing $\rho_1^{(n)}$ from $\rho_1^{(n)}$ can thus be 
expressed
as an average over the optimal mutual information of the subchannels:
\beq I_M = \sum_{j=1}^{2^n/2} q_j I_2(p_j),\label{Imutual}
\eeq where $q_j=\Tr
B_0^{[j]}=\Tr B_1^{[j]}$ is the probability of choosing the $j$'th
subchannel.
The BSC error probability $p_j$ for the
$j$'th subchannel depends on the subchannel's 2x2 renormalized 
density matrices $\hat{B}^{[j]}_p=B^{[j]}_p/q_j$, 
and is easily calculated once the optimal measurement is found.    
For each
subchannel the $q_j$ and renormalized 2x2 matrices look like
\beq
q_j=c^{2(n-k)} s^{2k} + c^{2k} s^{2(n-k)} \label{qj}
\eeq
and
\beq
\hat{B}_p^{[j]}=
\left(\begin{array}{cc}   
\frac{c^{2(n-k)} s^{2k}} 
                        {c^{2(n-k)} s^{2k} + c^{2k} s^{2(n-k)}} &
\frac{\pm c^n s^n} 
                  {c^{2(n-k)} s^{2k} + c^{2k} s^{2(n-k)})} \\
\vspace{0.2cm} 
\frac{\pm c^n s^n} 
                  {c^{2(n-k)} s^{2k} + c^{2k} s^{2(n-k)})} &
\frac{c^{2k} s^{2(n-k)}} 
                        {c^{2(n-k)} s^{2k} + c^{2k} s^{2(n-k)}} 
        \end{array}\right) \  \label{blocks} \ .
\eeq

In our previous example of $n=2$ the matrices are put in a block 
diagonal form:
\bara \rho_0^{(2)} = 
 \left(\begin{array}{cccc}    c^4   & c^2 s^2 &    0    &    0    \\ 
                            c^2 s^2 &   s^4   &    0    &    0    \\ 
                               0    &    0    & c^2 s^2 & c^2 s^2 \\ 
                               0    &    0    & c^2 s^2 & c^2 s^2 \\ 
  \end{array}\right) \ ; 
      \rho_1^{(2)} = 
 \left(\begin{array}{cccc}    c^4   &-c^2 s^2 &    0    &    0    \\ 
                           -c^2 s^2 &   s^4   &    0    &    0    \\ 
                               0    &    0    & c^2 s^2 &-c^2 s^2 \\ 
                               0    &    0    &-c^2 s^2 & c^2 s^2 \\
  \end{array}\right) \ , 
\ear
so that 
$q_{j=1} = c^4 + s^4$; $q_{j=2} = 2 c^2 s^2$; and 
\beq\label{bnormal} \hat{B}_p^{[j=1]} = 
\left(\begin{array}{cc}   
               \frac{c^4}{c^4+s^4} & \pm\frac{c^2 s^2}{c^4+s^4} \\ 
                   \pm \frac{c^2 s^2}{c^4+s^4} & \frac{s^4}{c^4+s^4}
       \end{array}\right) \ ;
\quad  \hat{B}_p^{[j=2]} = 
  \left(\begin{array}{cc}   1/2 & \pm 1/2  \\ 
                        \pm 1/2 &  1/2
      \end{array}\right) \ . \label{ex_n=2} \eeq

\section{Optimal Information in a Parity Bit}\label{sec3}

Two pure states or two density matrices in $\cH2$ with equal 
determinants can always be written (in an appropriate basis) in 
the simple form
\beq\label{amatrix}  \rho_0 = \left(\begin{array}{cc}   
                      a_1   &  a_2   \\ 
                      a_2   &  a_3   
     \end{array} \right) \ ; \quad
       \rho_1 = \left(\begin{array}{cc}   
                      a_1   & -a_2   \\ 
                     -a_2   &  a_3   
     \end{array} \right) \eeq
with $a_i$ real positive numbers such that 
$\Tr \rho_p = a_1 + a_3 = 1$.
For the two pure states of equation \ref{pure-state}, say, 
for the polarization states of a photon,
it is easy to see (and can be formally proven \cite{Levitin,Fuchs})
that a standard measurement in an orthogonal basis symmetric 
to the two states 
optimizes the mutual information (and also minimizes the 
avarage error probability).
The angle between one basis vector 
and the polarization state is
$\frac{\pi}{4} \pm \alpha$. The measurement results in an error 
with probability
\beq  P_e = \sin^2(\frac{\pi}{4}-\alpha)  
= \frac{1 - \cos(\frac{\pi}{2} - 2\alpha)}{2}
= \frac{1 - \sin(2 \alpha)}{2} \label{one-part} \ ,  \eeq
and with the same error probability for both inputs, 
thus, leading to a binary symmetric channel (BSC).
The optimal information of such a channel is well known and is
\beq I_{{}_{BSC}} = I_2(P_e)  \ . \eeq
Note that the overlap of the two-states is $\cos(2 \alpha)$, thus, 
for two pure states in any dimension,
the optimal 
information $I_2(\frac{1- \sin(2 \alpha)}{2}) $ is a simple 
function of the overlap.
The density matrices of such pure states (equation \ref{rho^1}) 
can be written as 
$\rho_i = (\one + \sigma \cdot {\bf r_i} )/2$ with the $\sigma$ being
the Pauli matrices and ${\bf r} = (\pm \sin 2 \alpha, 0 , \cos 2
\alpha)$ being a three dimensional vector
which describes a spin direction.
Using this notation any density matrix is described by a point in a 
three dimensional unit ball, called the Bloch sphere.
The pure states are points on the surface
of that sphere (also called the Poincare sphere).
With the density matrix notation  
the optimal basis for distinguishing the states
is the $x$ basis (note that the angle between the basis vector 
and the state is doubled  
in this notation). 
The measurement of the two
projectors
\beq      A_{\rightarrow} =  1/2 \left(\begin{array}{cc}   
                      1   &  1  \\ 
                      1   &  1 
     \end{array} \right) \quad \quad {\rm and} \quad \quad  
       A_{\leftarrow} =   1/2 \left(\begin{array}{cc}   
                      1   &  -1  \\ 
                     -1   &   1 
     \end{array} \right) \eeq
yields 
\beq P_e = \Tr \rho_1   A_{\rightarrow}  = \frac{1}{2} - a_2  
\label{Err-Prob} \ , \eeq
which recovers the result of equation (\ref{one-part}) in case of 
pure states
of equation (\ref{rho^1}).
However, the treatment of density matrices is more general
and this is the optimal measurement also 
in the case of non-pure states with equal 
determinants \cite{Levitin,Fuchs},  
when $\rho_i$ of equation (\ref{rho^1}) 
are replaced by $\rho_i^{dm}$ of 
equation (\ref{rhod_0}) and (\ref{rhod_1})
of section \ref{sec6},
and this case is also described by a BSC.  
The only difference between the matrices 
is that $\Det \rho_p = 0$ for pure states
and $ 0 \le \Det \rho_p \le \frac{1}{4} $ for density matrices. 

Instead of measuring the density matrices in the $x$ direction 
we perform the following unitary transformation  
on the density matrices 
\beq U =   1/\sqrt2 \left(\begin{array}{cc}   
                      1   &   1  \\ 
                      1   &   -1 
     \end{array} \right) \eeq
to obtain $\rho'=U \rho U^\dagger$ which is then 
measured in the $z$ basis.  
Note that the transformation transform the original $z$-basis 
to $x$-basis
(the motivation for this approach will be ubderstood when we 
discuss the 2x2 blocks of the parity matrices).
The new density matrices are 
\beq {\rho_0}' =\left(\begin{array}{cc}   
           \frac{1}{2} + a_2  &         \frac{a_1 - a_3}{2}    \\ 
           \frac{a_1 - a_3}{2}      &   \frac{1}{2} - a_2 
     \end{array} \right)    \ ; \   
     {\rho_1}' =\left(\begin{array}{cc}   
           \frac{1}{2} - a_2  &         \frac{a_1 - a_3}{2}    \\ 
           \frac{a_1 - a_3}{2}      &   \frac{1}{2} + a_2 
     \end{array} \right)  \label{dm_tag}   
\eeq
and their measurement yields the probability
$ \frac{1}{2} \pm a_2  $
to derive the correct (plus) and the wrong (minus) answers 
(as we obtained before), leading to 
optimal mutual information of 
\beq I_2(\frac{1}{2} - a_2)\ \label{BSC-info} ,\eeq
which depends only on $a_2$.
Note that the same information is obtained in case 
$a_1$ and $a_3$ are interchanged.

\medbreak

The naive way to derive information on a parity bit is to derive the
optimal information on each particle separately and calculate the
information on the parity bit.  We call this {\em individual} or {\em
single-particle} measurement.  It is the best Bob can do in case he
has no quantum memory in which to keep the particles (which, usually
arrive one at a time) or he has no ability to perform more advanced
joint measurements.  The optimal error-probability for each particle
is $ r \equiv P_e^{(1)} = \frac{1 - \sin 2 \alpha}{2} $.  The
probability of deriving the wrong parity bit is equal to the
probability of having an odd number of errors on the individual
particles
\bara  P_e^{(n)}  =
   \sum_{j = odd}^n {n \choose j} r^j (1-r)^{n-j} \ .\nonumber \ear
To perform the sum over only odd $j$ we use the formulas
\bara  (p+q)^n = 
\sum_{j = 0}^n {n \choose j} p^{n-j} q^j \ \ {\rm and}\
\ (p-q)^n = 
\sum_{j = 0}^n {n \choose j} p^{n-j} (-q)^j \ ,\nonumber \ear
to derive 
\beq 
\sum_{j = odd}^n {n \choose j} p^{n-j} q^j = 
\frac{(p+q)^n - (p-q)^n}{2} 
\ . \eeq
Assigning $q=r$ and $p=1-r$ we get
\beq   P_e^{(n)}  = \sum_{j = odd}^n {n \choose j} r^j (1-r)^{n-j}
= \frac{1^n - (1 - 2r)^n}{2} = \frac{1}{2} - \frac{(1-2r)^n}{2}  
       \label{n-part} \ . \eeq
The mutual information $I_S$ in this single-particle measurement is
\beq  
I_S = I_2(P_e^{(n)}) = I_2 \left(
\frac{1}{2} - \frac{(\sin 2\alpha)^n}{2} \right)
\eeq
using $ r = \frac {1-\sin(2\alpha)}{2}$.

A lot of useless side-information is also obtained (e.g., on the
individual bits). This fact indicates that Bob might be able to do
much better by concentrating on deriving only useful information.
The optimal measurement for finding mutual information on the 
parity bit
is not a single-particle measurement, but is instead a measurement on
the full $2^n$-dimensional Hilbert space of the system.  In general,
optimizing over all possible measurement is a very difficult task
unless the two density matrices in $\cH{2^n}$ are pure states.
However, in the preceding section we have shown how to reduce the
problem to that of distinguishing the 2x2 blocks of our
block-diagnoral parity matricies.  We now have only to apply the
optimal single-particle measurement to the 2x2 $\hat{B}^{[j]}$'s of
equation~\ref{blocks} and use the result in equation~\ref{Imutual}.

The error probability (equation~\ref{Err-Prob}) for distinguising 
the $\hat{B}^{[j]}$'s is seen to be:
\beq p_{j} = \frac{1}{2} - \frac{c^n s^n}{ c^{2(n-k)} s^{2k} 
+  c^{2k} s^{2(n-k)} }
\ ,\label{pj}  \eeq
from which the information $I_2(p_j)$ in each channel is obtained.

Plugging the error probability $p_j$ (equation \ref{pj}) 
and the probability of choosing the j'th subchannel $q_j$ 
(equation~\ref{qj})
into \ref{Imutual}, the 
optimal information on the parity bit is now:
\beq I_M = \sum_{j=1}^{2^n/2} 
\big( c^{2(n-k)} s^{2k} + c^{2k} s^{2(n-k)} \big)
\   I_2\left( \frac{1}{2} - \frac{c^n s^n}
   {c^{2(n-k)} s^{2k} + c^{2k} s^{2(n-k)}}  \right) \ . 
\label{I-mutual} \eeq

In the simple case of orthogonal states ($\alpha = \frac{\pi}{4}$) 
all these density matrices are the same and we get
$q_j = \left(\frac{1}{2}\right)^{n-1}$, $\ p_j = 0$ and $I_M = 1$
as expected.
\begin{itemize}
\item A brief remark is in order at this stage.
The transformation to the $x$ basis for each 2 by 2 matrix, 
$\hat{B}_p^{(n,k)}$ is actually a transformation from a product basis
to a fully entangled basis of the $n$ particles.
That basis is a generalization of the Bell basis of \cite{BMR}.  
\beq  {1 \choose 0}_1 
      {1 \choose 0}_2 
       \ldots       
      {1 \choose 0}_{n-1}
      {1 \choose 0}_n  \pm  
      {0 \choose 1}_1 
      {0 \choose 1}_2 
       \ldots       
      {0 \choose 1}_{n-1}
      {0 \choose 1}_n 
 \ ; \nonumber \eeq   
\beq  {1 \choose 0}_1 
      {1 \choose 0}_2 
       \ldots       
      {1 \choose 0}_{n-1}
      {0 \choose 1}_n  \pm  
      {0 \choose 1}_1 
      {0 \choose 1}_2 
       \ldots       
      {0 \choose 1}_{n-1}
      {1 \choose 0}_n 
\eeq   
etc.
The Bell basis for two particles is frequently used and its basis 
contains the EPR singlet state and other three orthogonal fully 
entangled states.
\end{itemize}

For large $n$, the number of blocks is exponentially large
and performing the summation required in equation \ref{I-mutual} 
is impractical,
since all the $2^{n-1}$ matrices must be taken into account.
However, that problem can be simplfied by realizing that all blocks
with a given $k$, as well as all blocks with $k$ and $n-k$ 
interchanged,
contribute the same information to the total.  This is easily seen
in equation~\ref{I-mutual} where both the weight and the argument
of $I_2$ are symmetric in $k$ and $n-k$.
The optimal mutual information for even $n$ is then
\beq  
I_M^{even} = \sum_{k=0}^{\frac{n}{2}-1} {n \choose k} q_k I_2(p_k) + 
\frac{1}{2} 
{n \choose \frac{n}{2}} q_{\frac{n}{2}} I_2(p_{\frac{n}{2}})\ , 
\label{Ieven} \eeq
and for odd $n$
\beq I_{M}^{odd} =  \sum_{k=0}^{\frac{n-1}{2}} 
{n \choose k} q_k I_2(p_k) \ .
\label{Iodd}  \eeq

As an example we calculate $I_M$ for $n=2$ (of course, 
the counting argument is not needed in that case).  This particular 
result complements the result in \cite{HP} where the deterministic 
information 
of such a system is considered (see also Section \ref{sec5}).  
In the new basis (\ref{dm_tag}) the density matrices of 
(equation~\ref{ex_n=2})
become
\bara  \hat{B}_0^{'(n=2,k=0)} =
  \left(\begin{array}{cc}  1/2 + \frac{c^2 s^2}{c^4 + s^4}  & 
                         \frac{1}{2}\frac{c^4-s^4}{c^4+s^4}  \\ 
        \frac{1}{2}\frac{c^4-s^4}{c^4+s^4} &      
                           1/2 - \frac{c^2 s^2}{c^4 + s^4}   
        \end{array}\right) \ ; \quad
 \hat{B}_1^{'(n=2,k=0)} =
  \left(\begin{array}{cc}  1/2 - \frac{c^2 s^2}{c^4 + s^4}  &
 \frac{1}{2}\frac{c^4-s^4}{c^4+s^4}  \\ 
        \frac{1}{2}\frac{c^4-s^4}{c^4+s^4} &      
                           1/2 + \frac{c^2 s^2}{c^4 + s^4}   
        \end{array}\right)
\nonumber \ear and 
\beq   \hat{B}_0^{'(n=2,k=1)} =
  \left(\begin{array}{cc}                1    &     0 \\ 
                                         0    &     0 
      \end{array}\right) \ ; 
\quad  \hat{B}_1^{'(n=2,k=1)} = 
  \left(\begin{array}{cc}                0    &     0 \\ 
                                         0    &     1 
      \end{array}\right) \ . \nonumber \eeq
We use the notation $S = 2sc = \sin 2\alpha$; 
$C = c^2 - s^2 = \cos 2 \alpha$
(hence,  $c^4 - s^4 = C$ and  $ c^4 + s^4 = 
\frac{1+C^2}{2}$)
to obtain
$q_1 = 2c^2 s^2 = \frac{S}{2}$, $p_1 = 0$,    
$q_0 = \frac{1}{2} (1 + C^2)$ and  
$p_0 = \frac{C^2}{1+C^2}$ (the $q_j$s were obtained in the previous
section).
The mutual information of the parity of two bits is obtained using
equation (\ref{Imutual})
\bara I_M &=& q_0 I_2(p_0) +  q_1 I_2(p_1)  \nonumber \\ &=&  
      \frac{1}{2} (1 + C^2) 
 I_2 \left(\frac{C^2}{1+C^2}\right) + \frac{S^2}{2} \ . \ear

\section{Information on the Parity Bit of Almost Fully Overlapping 
States}\label{sec4} 

The case of almost fully overlapping states is extremely important to
the analysis of eavesdropping attacks on any quantum key distribution
scheme as will be discussed in Section \ref{secd}.  In this case the
angle $\alpha$ is small so $s\equiv\sin \alpha \simeq \alpha$ and
$c\equiv\cos \simeq 1 - \frac{\alpha^2}{2}$.  To observe the 
advantage of the joint measurement, let us first calculate the 
optimal information obtained by individual measurements.  In that 
case,equations (\ref{n-part} and (\ref{one-part}) yield
\beq P_e^{(n)}  = \frac{1}{2} - \frac{(2 \alpha)^n}{2} \ .\eeq
For small $\eta$ the logarithmic function is approximated by
\beq 
\log(\frac{1}{2} \pm \eta ) = \frac{\ln(\frac{1}{2} \pm \eta)}{\ln 2}
\approx
 -1 \pm \frac{2}{\ln 2} \eta - \frac{2}{\ln 2} \eta^2 \ , \eeq
from which the mutual information  
\bara I_2 (\frac{1}{2} - \eta) &=& 1 - H(\frac{1}{2} - \eta) = 1 + 
                (\frac{1}{2} + \eta)  \log(\frac{1}{2} + \eta) +
       (\frac{1}{2} - \eta)  \log(\frac{1}{2} - \eta) \nonumber \\ 
           &\approx& \frac{2}{\ln 2} \eta^2 \label{Iapprox}  \ear
is obtained.  Using this result and assigning $\eta = (2\alpha)^n/2$,
the information (to first order)
obtained by the optimal single-particle measurement is
\beq I_S = \frac{2}{\ln 2} \frac{(2 \alpha)^{2n}}{4}
= \frac{(2 \alpha)^{2n}}{2 \ln 2}.  \label{Iindiv} \eeq

We use the same approximations and equations (\ref{pj}) 
and (\ref{qj}) to calculate the leading terms in the 
optimal mutual
information \ref{Ieven} and \ref{Iodd}.
For $k = \frac{n}{2}$ ($n$ even) we get $p_k = 0$ 
(regardless of the small angle) and 
\beq I_2(p_{\frac{n}{2}}) = 1 \ . \eeq
For $k < \frac{n}{2}\  $ we get 
$p_k \approx \frac{1}{2} - \frac{s^n}{s^{2k}} \approx
 \frac{1}{2} - \alpha^{n-2k} $ 
which yields 
(using equation~\ref{Iapprox} 
with $\eta = \alpha^{n-2k}$)
\beq  I_2(p_k) \approx
 \frac{2}{\ln 2} \alpha^{2n - 4k} \ . \eeq 
The coefficient $q_k = \alpha^{2k}$ for $k< \frac{n}{2}$ and
$q_k = 2 \alpha^{2k}$ for $k= \frac{n}{2}$, 
so that 
\beq q_k I_2(p_k) \approx
 \frac{2}{\ln 2} \alpha^{2(n-k)} \  \eeq
for $k < \frac{n}{2}$, and  
\beq q_k I_2(p_k) \approx  2 \alpha^n \  \eeq
for $k = \frac{n}{2}$. 
The dominant terms are those with the largest $k$, that is, $k$ 
closest to $\frac{n}{2}$.
The next terms are smaller by two orders in $\alpha$.
The number of density matrices with these $k$'s are also the largest
(up to a factor of 2 in case of even $n$). 
Therefore, the terms
$k= \frac{n}{2}$ for even $n$ and $k= \frac{n-1}{2}$ for odd $n$
are the dominant terms in the final expression.
Thus, for almost fully overlapping states, the mutual information is
\bara 
I_M^{even} \approx \frac{1}{2} {n \choose \frac{n}{2}  } 2 \alpha^n 
 = { n \choose \frac{n}{2} } \alpha^n \nonumber \ear 
for even $n$,
and 
\beq 
I_M^{odd} \approx {n \choose \frac{n-1}{2} }
         \frac{2}{\ln 2} \alpha^{n+1}  \label{Iopt} \eeq
for odd $n$.
 
These expressions can be further simplified.
The number of density matrices of any type is bounded (for large $n$)
using Stirling formula (see \cite{MS} in the chapter on 
Reed-Solomon codes)  
\beq  
{ n \choose k} < \frac{2^{nH(k/n)}}{\sqrt{2\pi (k/n)(1-k/n) n}} \ .
\eeq
For $k$ near $\frac{n}{2}$,  
$\eta \equiv \frac{1}{2} - \frac{k}{n}$ is small, and the 
standard approximation (\ref{Iapprox}):
$H \approx 
1 - O\big(\eta^2\big) = 
1 - O\left((\frac{1}{2} - k/n)^2 \right) < 1 $
is used to derive 
${ n \choose k} < \frac{2^n}{\sqrt{2\pi (k/n)(1-k/n) n}}$.
Using also $k/n(1-k/n) \approx \frac{1}{4} -  
  \eta^2$, we derive 
\beq
{ n \choose k} < \frac{2^n}{\sqrt {\frac{\pi}{2} n}} (1+O(\eta^2))\ .
\eeq

Thus the leading term in $I_M$ is
\beq I_M < \frac{2^n}{\sqrt{\frac{\pi}{2} n}} \alpha^n = 
             (2\alpha)^n /\sqrt{\frac{\pi}{2} n}  \eeq
for even $n$
and 
\beq 
I_M < \frac{2^n}{\sqrt{\frac{\pi}{2} n}} 
         \frac{2}{\ln 2} \alpha^{n+1} = 
         \frac{2}{\ln 2} \alpha
                 (2 \alpha)^n  /\sqrt{\frac{\pi}{2} n}  
             <   (2 \alpha)^n  /\sqrt{\frac{\pi}{2} n}  \eeq
for odd $n$ (using $\alpha < \ln2 / 2$). 
We see that we could keep a better bound for odd $n$ but for
simplicity we consider the same bound for both even and odd $n$'s.

We can now compare the optimal information $I_M$ from a joint
measurement on all $n$ particles to the optimal information $I_S$
from separate measurements (cf. eq.~\ref{Iindiv}): 
\beq\begin{array}{ccl} I_M & = & O(1)\times (2\alpha)^n/\sqrt{n} \\
I_S & = & O(1)\times (2\alpha)^{2n}. \\
\end{array}\eeq Since $\alpha$ is a small number (corresponding to
highly overlapping signal states), the joint measurement is 
superior tothe individual measurement by a factor of 
$ O\big(\ (2\alpha)^{n}\big)$. 
However, it is only superior by a
polynomial factor, since 
\beq I_M \approx (I_S)^2\ . \eeq

\section{Deterministic Information on the Parity Bit}\label{sec5}

For a single particle Bob can perform a different kind of individual
measurement which is not optimal in terms of avarage mutual 
information but is sometimes very useful \cite{Ben92,Per93}.
It yields either a conclusive result about the value
of that bit or an inconclusive one, and Bob will know which of the
types of information he has succeeded in obtaining.
Such a measurement corresponds to a binary erasure channel
\cite{Per93,HP,EHPP}.
With probability $p_?$ of an inconclusive result, 
the mutual information is 
$I_{p_?} = 1 - p_?$.
The minimal probability for an inconclusive result is
$\cos 2\alpha$  leading to 
$I_{p_?} = 1-\cos 2\alpha$ \cite{Per93}. This result is obtained
by performing a generalized measurement (Positive Operator Value 
Measure \cite{Per93,Helstrom,JP-DL})
on the system or a standard measurement performed on a larger 
system which contains the system and an auxilary particle 
\cite{Per93,IvPe}. Note that this results in less mutual information 
than the optimal measurement for one-particle mutual information.
If Bob uses this type of measurement on each particle separately 
his deterministic single-particle information about
the parity bit is $(1-\cos 2\alpha)^n$.

We now use the block-diagonal density matrices derived in 
section~\ref{sec2}
to derive also the optimal {\em deterministic} 
information on the parity bit.  We note that each of the 2x2 blocks 
in the block-diagonal
density matrices is the density matrix of a pure state, so we
may replace the optimal measurement in each subchannel with 
the optimal deterministic measurement and proceed as before.
The total optimal deterministic 
information is easily calculated by replacing $I_2(p_k)$ in 
\ref{Ieven} and \ref{Iodd} by 
$I(p_{?_k}) = 1 - p_{?_k}$.
To find the minimal 
$P_{?_k}$ we write each of the normalized density 
matrices $\hat{B}_p^{(n,k)}$ as pure states
with some angle $\gamma$: 
\beq 
{\cos\gamma \choose \sin\gamma} \quad \quad {\rm and} \quad \quad
{\cos \gamma \choose -\sin \gamma} \ .  \eeq
Comparing with equation (\ref{blocks})
\beq p_? = \cos(2 \gamma) = 
\cos^2\gamma - \sin^2\gamma = 
\frac{c^{2(n-k)} s^{2k} - c^{2k} s^{2(n-k)}} 
                  { c^{2(n-k)} s^{2k} + c^{2k} s^{2(n-k)}} \ , \eeq
hence
\beq I(p_{?_k})  = 1 - 
\frac{c^{2(n-k)} s^{2k} - c^{2k} s^{2(n-k)}} 
                  {c^{2(n-k)} s^{2k} + c^{2k} s^{2(n-k)}} \ .\eeq
The total information is 
\beq  
I_{D}^{even} = \sum_{k=0}^{\frac{n}{2}-1} {n \choose k} 
q_k I(p_{?_k}) 
+ \frac{1}{2} {n \choose \frac{n}{2}} q_{\frac{n}{2}} 
I(p_{?_{n/2}})\ , 
\label{Ideteven} \eeq
for even $n$, and
\beq 
I_{D}^{odd} = 
\sum_{k=0}^{\frac{n-1}{2}} {n \choose k} q_k I(p_{?_k}) 
\label{Idetodd}  \eeq
for odd $n$.

For $n=2$ we 
recover a result previously obtained by Huttner and Peres 
\cite{HP} by performing the optimal POVM on the first pair of
density matrices of 
equation (\ref{ex_n=2}), 
and a measurement in the entangled basis (as before) on the second.
The probability of an inconclusive result is
$\cos^2\gamma - \sin^2 \gamma = \frac{c^4 - s^4}{c^4 + s^4} 
= \frac{2C}{1+C^2}$, 
hence the optimal deterministic information is
$1 -  \frac{2C}{1+C^2}$, 
leading to the total deterministic information
\beq I_D = q_1 I_D(p_?) + q_2 I_2(p_2) =  
      \frac{1}{2} (1 + C^2) 
          (1 - 2\frac{C}{1+C^2}) + \frac{S^2}{2} = 1 - C \ , \eeq
which is exactly the result obtained by Huttner and Peres (note, 
however,
that they used an angle which is 
$\frac{\pi}{4}-\alpha$ hence derived $1-S$ for the 
deterministic information).

For almost overlapping states (small $\alpha$) the dominant terms
are still the same as in the case of optimal information.
The term $q_k$ is as before
and the information in each port is \beq I(p_{?_k})=1 \eeq 
for $k = \frac{n}{2}$ and 
\beq I(p_{?_k}) = 1 - \frac{c^{2n-4k} - s^{2n-4k}}{
                         c^{2n-4k} + s^{2n-4k}} =
           1 - (1 - \alpha^{2n-4k})^2  = 2 \alpha^{2n-4k}         
   \eeq
for $k<\frac{n}{2}$.
Taking into consideration only the dominant term we get 
\bara I_D^{even} 
\approx  {n \choose \frac{n}{2}  }   \alpha^n
\nonumber \ear 
for even $n$ which is the same as the optimal information, 
and 
\beq 
I_D^{odd} \approx {n \choose \frac{n-1}{2} }
         2 \alpha^{n+1}   \eeq
for odd $n$ which is smaller 
than the optimal mutual information by a factor of $\frac{1}{\ln 2}$.

\section{Parity Bit for Density Matrices}\label{sec6}

The previous disussion assumed that $\rho_p^{(1)}$ are pure states.
The generalization to the case of density matrices with equal 
determinants is straightforward.
Let the bit `0' and the bit `1' be represented by  
\beq \rho_0^{dm} = 
  \left(\begin{array}{cc}   c^2 &  sc-r  \\  sc-r &  s^2     
  \end{array}\right),\label{rhod_0} \eeq
and 
\beq \rho_1^{dm} = 
  \left(\begin{array}{cc}   c^2 &  -(sc-r)  \\  -(sc-r) &  s^2     
  \end{array}\right),\label{rhod_1} \eeq
(with $s=\sin \alpha$ etc., and $r<sc$)
which contains the most general density matrices of the desired type.
On the Poincare sphere 
these density matrices have the same $z$ components
as the previously written pure states but smaller $x$ components
(hence smaller angle $\alpha'$).
We could choose other ways of representing these density matrices,
e.g., with similar $x$ components and smaller $z$ components.
Such representations were more appropriate for comparison with pure
states (since they yield the same mutual information for a single
particle) but less convenient for showing that the previous result
is easily generalized.

Clearly 
\beq \rho^{(1)_{dm}} = \frac{1}{2} (\rho_0^{(1)} + \rho_1^{(1)}) = 
  \left(\begin{array}{cc}   c^2 &  0  \\  0 &  s^2     
  \end{array}\right),\label{rhod} \eeq
\beq \Delta^{(1)_{dm}} = \frac{1}{2} (\rho_0^{(1)} - \rho_1^{(1)}) = 
  \left(\begin{array}{cc}   0 &  sc-r  \\  sc-r &  0     
  \end{array}\right),\label{Deltad} \eeq
The total density matrix doesn't change and the difference density 
matrix
has terms $(sc - r)^n$ instead of $(sc)^n$.
Reorganizing the basis vectors we again get the block diagonal 
matrices
where each of the 2 by 2 matrices has the form 
\beq   B_p^{(n,k)} = \left(\begin{array}{cc}   
         c^{2(n-k)} s^{2k}    &  \pm (cs-r)^n             \\ 
         \pm (cs-r)^n          &  c^{2k} s^{2(n-k)} 
  \end{array}\right). \label{Bpnk-dm} \eeq
When normalized, these density matrices have the form of 
equation~\ref{amatrix} and are optimaly distinguished by measuring 
them in the $x$ direction. Transforming to the $x$ basis as before 
we get the same 
\beq q_k = 
       c^{2(n-k)} s^{2k} 
                 +  c^{2k} s^{2(n-k)} \eeq
as before, and 
\beq  p_k = \frac{1}{2} - \frac{(cs-r)^n}{ 
                                          c^{2(n-k)} s^{2k} 
                                    +  c^{2k} s^{2(n-k)}} \ . \eeq
The total information can now be calculated as before by assigning 
these $p_k$ and $q_k$ into equations (\ref{Iodd} and \ref{Ieven}).
Thus, the case of mixed states is also analytically solved for any 
number of bits, and the influence of mixing on the optimal mutual 
information is through the $p_k$s.

The case of $\alpha = \frac{\pi}{4}$ (when the matricies commute) is 
of special interest\footnote{This case, including the X-shape of 
the parity
matrices, was solved independently by D. Mayers \protect\cite{DM}.}
due to its imporance to quantum bit committment \cite{BCJL}.
Equation (\ref{Bpnk-dm}) yields
\beq   B_p^{(n,k)} = \left(\begin{array}{cc}   
      (\frac{1}{2})^n       &  \pm (\frac{1}{2}-r)^n             \\ 
         \pm (\frac{1}{2}-r)^n          &  (\frac{1}{2})^n   
  \end{array}\right)  \eeq
indepent of $k$.
Normalizing and transforming to the entangled basis we get
$p = \frac{1}{2} - \frac{1}{2}(1-2r)^n$ and $q=(\frac{1}{2})^{n-1}$.
There are $2^{n-1}$ matrices like that so 
\beq I_M = 2^{n-1}
(\frac{1}{2})^{n-1}
I_2\left(\frac{1}{2} - \frac{1}{2}(1-2r)^n\right) = 
I_2\left(\frac{1}{2} - \frac{1}{2}(1-2r)^n\right) \ . \eeq
However, in this case each particle actually carries a classical
information hence the collective measurement cannot improve the 
derived information.
Indeed, the one-particle density matrices  yields probability $r$ of 
deriving an error, leading to total error probability of equation 
(\ref{n-part}),
$ P_e^{(n)} =
              \frac{1}{2} - \frac{(1-2r)^n}{2} $,
and total information $I_S
=I_2\left(\frac{1}{2} - \frac{1}{2}(1-2r)^n\right) $
as expected.
Note that the mixing by itself induces an exponential decay of the 
amount of
information, a fact used in \cite{BCJL}.

Calculating the optimal information for small $\alpha$ and any 
$r$ is possible but complicated.
Another alternative which is much simpler is to find a bound on the 
optimal
information using pure states with the same angle, $\alpha'$, 
on the Poincare sphere, using
\beq 
\tan 2 \alpha' = \frac{\sin 2\alpha - 2 r}{\cos 2\alpha} \ ,\eeq  
or using an alternative form for the mixed 
states \ref{rhod_0} and \ref{rhod_1}.

\section{Implications}\label{secd}

Protocols in quantum cryptography use parity bits.
Quantum bit commitment \cite{BCJL},
quantum oblivious transfer \cite{BBCS} and quantum key distribution
\cite{BBBSS} protocols use parities of publicly announced subsets 
of the
transmitted bits for both error-correction and privacy 
amplification (PA).
When used for error-correction, subset parities are publicly 
announced 
in order to identify errors and correct them, and this is a crucial 
step
in real channels since it could leak information to Eve. When used 
for PA
\cite{BBR,BBCM} (say, to derive one final bit) a subset parity
is agreed to be the final secret bit, and this technique is used 
to limit the adversary's information to an exponentially small 
fraction of a bit.
PA is effective when particles are not measured together 
(see discussions in \cite{BBCM} and in \cite{DMLS}), and presumably 
also if all measurements are completed
before the specification of the subsets used in PA is publicly 
announced.
But it is still an open question whether it is effective also when 
the 
adversary can use this specification to {\em choose} her attack.
Our result provides the optimal measurement which can be done to 
find a 
parity bit and therefore is crucial for such analysis.
In particular cases, when almost fully overlapping states are used,
we proved two complementary results regarding that optimal 
measurement:
\begin{itemize}
\item The optimal information is
much larger than the one obtained be measuring 
each bit separately.
\item The optimal measurement still yields
exponentially small information. Thus we proved
an {\em effectiveness result}: classical PA techniques are effective
against {\em any} quantum measurement.
\end{itemize}
The discussion so far treats an imaginary scenario which is very 
general 
but is not good as a cryptographic protocol.

Realistic protocols are very complicated (e.g., due to the use of 
error
correction), hence, are more difficult to analyze. However,
to emphasize the importance of the ``effectiveness result'' just 
mentioned
let us consider a different scenario which is common in quantum key
distribution schemes:
Alice and Bob are the legitimate users who try to establish a secret
key.
They use 
any binary scheme and Alice sends $n$ particles through a noisy
channel to Bob.  
An adversary, Eve, is trying to learn information on their key.
She gets the particles one at a time, 
interacts with each one of them weakly, and send it forward to Bob.
She must interact weakly with all particles 
if she wants to induce only small error rate
(or else she could attack strongly only few of the particles but 
PA is already proven effective against that type of attacks). 
The classical PA is also effective if Eve
cannot use its specification to attack all bits together, but in 
reality
she can do it if she has a quantum memory.
Although the specification is announced after the transmission is 
over, 
Eve can keep information in the quantum state of a system which has
interacted with all the transmitted particles, and use it after all
the specification is announced.
Such attack has never been analyzed in the case
of real (noisy) channels and devices.\footnote{
The case of error-free channels is completely solved, a consequence 
of 
\cite{Yao,Mayers}. The case of error-free devices (but real channels)
can be solved due to the 
possibility of purifying singlets \cite{purification,purification2}.}
The ``effectiveness result'' might allow one to prove security
against a ``collective'' attack (first described by E.~Biham 
and T.~Mor)
in which Eve learns the PA specification and the error-correction 
data
and uses them to choose the optimal measurement.
In that atack Eve attaches a separate probe to each particle via
transluscent attack (defined in \cite{EHPP}), using a quantum memory
to keep
the probes for a later time, and measuring them {\em together} after
receiving all relevant data.
Eve must attack weakly since she doesn't want to induce large error 
rate,
so she obtains (for each transmitted particle)
a probe with two almost overlapping pure states (or density 
matrices).
Hence, the ``effectiveness result'' can be used to prove that her 
information on the final string is exponentially small in the length 
of the initial string.

\bigbreak

\section*{Acknowledgments}
Thanks are due to Asher Peres for suggesting the 
problem and for many helpful discussions.
We also thank E.~Biham, S.~Braunstein, C.~Fuchs, and
B.~Wootters for helpful discussions.
We also thank the Institute for Scientific Interchange in Torino, 
Italy
for allowing us to meet and collaberate, 
and Gilles Brassard and the Universit\'{e} de Montr\'{e}al for 
hosting another productive meeting.
Some of these results were derived independently by Dominic Mayers.


\begin{thebibliography}{99}
\bibitem{Hol} A. S. Holevo, {\it Probl. Inform. Transmission} 
{\bf 9}, 
110 (1973).
\bibitem{Davies} E. B. Davies, {\it IEEE Trans. Inform. Theory} 
{\bf IT-24},
596 (1978).
\bibitem{Helstrom} C. W.  Helstrom, {\it Quantum Detection and 
Estimation
Theory }, Academic Press, New York (1976).
\bibitem{Per93} A. Peres, {\it Quantum Theory: Concepts and
Methods\/}, Kluwer, Dordrecht (1993), Chapt.~9.
\bibitem{Levitin} L. B. Levitin {\it Proc. of the workshop on 
Phys. of Comput.: PhysComp 92, IEEE}, 210 (1993). 
\bibitem{Fuchs} C. A. Fuchs and C. M. Caves,
{\it Phys. Rev. Lett.} 
{\bf 73}, 3047 (1994).
\bibitem{BB84} C.~H.~Bennett and G.~Brassard, in {\em Proceedings 
of IEEE
International Conference on Computers, Systems and Signal Processing,
Bangalore, India} (IEEE, New York, 1984) p.~175.
\bibitem{Eke91} A. K. Ekert, {\it Phys. Rev. Lett.}  {\bf 67},
661 (1991); C. H. Bennett, G. Brassard and N. D. Mermin, 
{\it Phys. Rev. Lett.} 
{\bf 68}, 557 (1992).
\bibitem{Ben92} C. H. Bennett, {\it Phys. Rev. Lett.}
{\bf 68}, 3121 (1992).
\bibitem{BCJL} G. Brassard, C. Crepeau, R. Jozsa and D. Langlois,
{\it Proc. 34${}^{th}$ Annual Symposium on Foundation of Computer 
Science, 
IEEE Press.}, 362 (1993).
\bibitem{BBBSS} C.H.~Bennett, F.~Bessette, G.~Brassard, L.~Salvail 
and J.~Smolin, J. Cryptology {\bf 5}, 3 (1992).
\bibitem{BBR}
C.H.~Bennett, G.~Brassard and J-M.~Robert,
{\em Siam J.\ Comput}, {\bf 17}, 210 (1988).
\bibitem{BBCM} C. H. Bennett, G. Brassard, C. Crepeau and U. Maurer,
{\it Privacy Amplification}, IEEE Trans. Info. Theo. {\bf 41}, 
1915 (1995).
\bibitem{HP} B. Huttner and A. Peres, 
{\it J. Mod. Opt.} {\bf 41}, 2397 (1994). 
\bibitem{BMR} S. L. Braunstein, A. Mann and M. Revzen,  
{\it Phys. Rev. Lett.} 
{\bf 68}, 3259 (1992).
\bibitem{MS} F. J. MacWilliam and N. J. A. Sloane, {\it The Theory 
of Error-Correcting Codes}, North Holand, 1977. 
\bibitem{EHPP} A.K.~Ekert, B.~Huttner, G.M.~Palma and A.~Peres, 
{\it Phys. Rev. A.} {\bf 50}, 1047 (1994).
\bibitem{JP-DL} J. M. Jauch and C. Piron, {\it Helv. Phys. Acta.} 
{\bf 40}, 559 (1967); E. B. Davies and J. T. Lewis, {\it Com. Math. 
Phys. } {\bf 17}, 239 (1970).
\bibitem{IvPe} I.~D.~Ivanovic, {\it Phys. Lett. A} 
{\bf 123}, 257 (1987); A.~Peres, 
{\it Phys. Lett. A} {\bf 128}, 19 (1988). 
\bibitem{DM} D.~Mayers, talk at ISI Quantum Computation Workshop, 
June 1995.
\bibitem{BBCS} C. H. Bennett, G. Brassard C. Crepeau and M. H. 
Skubiszewska,
{\it Proceedings of Crypto '91. LNCS} {\bf 576}, 351 (1992).
\bibitem{DMLS} D. Mayers and L. Salvail, {\it PhysComp 94}, 69,
Dalas (1994).
\bibitem{Yao} A.~Yao, ACM Symposium on Theory of Computing, Las 
Vegas (1995).
\bibitem{Mayers} D. Mayers, {\it Proceedings of Crypto '95. LNCS} 
{\bf 963},
124 (1995).
\bibitem{purification} C.H.~Bennett, G.~Brassard, S.~Popescu,
B~Schumacher,J.A.~Smolin, and W.K.~Wootters,``Purification of Noisy 
Entanglement and Faithful Teleportation via Noisy Channels'', 
{\em Phys. Rev. Lett.} {\bf 76}, 722 (1996).
\bibitem{purification2} A.~Ekert {\em et. al.}, ``Notes on quantum 
cryptography over noisy channels with quantum privacy 
amplification'', unpublished.

\end{thebibliography}
\end{document}